# Acoustic excitation and electrical detection of spin waves and spin currents in hypersonic bulk waves resonator with YIG/Pt system

N. I. Polzikova[a,*], S. G. Alekseev[a], V. A. Luzanov[b], A. O. Raevskiy[b]

[a]*Kotel'nikov Institute of Radio Engineering and Electronics of Russian Academy of Sciences,*

*Mokhovaya str. 11, build. 7, Moscow, 125009, Russia*

[b]*Fryazino branch Kotel'nikov Institute of Radio Engineering and Electronics of Russian Academy of Sciences*

*Vvedenskiy sq.,1, Fryazino, Moscow Region, 141190, Russia*

## Abstract

We report on the self-consisted semi-analytical theory of magnetoelastic excitation and electrical detection of spin waves and spin currents in hypersonic bulk acoustic waves resonator with ZnO-GGG-YIG/Pt layered structure. Electrical detection of acoustically driven spin waves occurs due to spin pumping from YIG to Pt and inverse spin Hall (ISHE) effect in Pt as well as due to electrical response of ZnO piezotransducer. The frequency-field dependences of the resonator frequencies and ISHE voltage $U_{ISHE}$ are correlated with experimental ones observed previously. Their fitting allows to determine some magnetic and magnetoelastic parameters of YIG. The analysis of the YIG film thickness influence on $U_{ISHE}$ gives the possibility to find the optimal thickness for maximal $U_{ISHE}$ value.

## 1.Introduction

In recent years, acoustically driven spin waves (ADSW) are of great interest in connection with the key objectives of next-generation spin-based technologies [1-14]. The piezoelectric generation of ADSW in composite magnetoelastic structures [8 - 14] is promising for use in low power consumption devices free from energy dissipation due to ohmic losses. In particular, acoustic spin pumping - the generation of spin-polarized electron currents from ADSW [7-9] - is promising for microwave spintronics and attracts much attention of the researchers.

The acoustic waves (AW) and spin waves (SW) coupling, caused by linear magnetostriction, is most significant under the condition of the phase synchronism, i.e. at magnetoelastic resonance (MER). As far as MER frequency generally lies in the gigahertz range [15] the generation of hypersonic AW is required. At present, it is considered that bulk AW are promising for applications at frequencies above 2.5-3 GHz. One of the ways to excite bulk AW with the frequencies up to 20 GHz is the use of a high overtone ($n \sim 10^2 \div 10^3$) bulk acoustic wave resonator (HBAR) [16]. Previously, in [13,14] we demonstrated the piezoelectric excitation of ADSW at 2 GHz by means of HBAR containing ferrimagnetic yttrium iron garnet (YIG) and piezoelectric zinc oxide (ZnO) films. Quite recently, the resonant acoustic spin pumping in HBAR containing YIG/ Pt system was proposed and implemented in our works [17, 18].

This paper presents a theoretical consideration for acoustic spin pumping in HBAR and detection of the ADSW through the inverse spin Hall effect (ISHE) in Pt. Accounting for the back action of ADSW in YIG on the elastic system in all layers of the structure (in non-magnetic layers through boundary conditions) makes it possible to determine and compare the frequency and magnetic field dependences of HBAR resonance frequencies, $f_n$, and dc ISHE voltage, $U_{ISHE}$. Comparison of these theoretical and experimental [18] dependences shows a qualitative agreement and allows us to determine a number of magnetic parameters of the YIG films. The calculation also shows that there is the optimal YIG film thickness for acoustic spin pumping efficiency, which may be an order of magnitude higher than observed previously by means of HBAR.

## 2. HBAR structure

In Fig.1 the HBAR structure is shown. It contains a gadolinium-gallium garnet (GGG) substrate *4* and two YIG films *3*, *5* on both sides of the substrate [19]. A bulk AW transducer consisting of a piezoelectric ZnO film *1*, sandwiched between thin-film Al electrodes *2*, is deposited on one side of the YIG-GGG-YIG structure. To excite the bulk AW propagating along the *x*-axis, the rf voltage $\tilde{U}(f)$ with frequency $f$ is applied across the transducer. A thin Pt strip *6* is attached to the YIG film *5* underneath the acoustic resonator aperture. Below we will use for the layer with the index $i = 1...6$ the notations $l^{(i)}$ for the thickness and $x_i$ for the coordinate of the lower surface. The external magnetic field **H** lies in the plane of the structure along the *z*-axis and magnetizes YIG films up to uniform saturation magnetization $\mathbf{M}_0 \| \mathbf{H}$.

It is assumed that ZnO film with an inclination of piezoelectric **c**-axis excites shear bulk AW polarized along the *z*-axis [20]. In YIG layers, this wave drives magnetization dynamics due to the magnetoelastic interaction. The AW and SW interaction results in the shift $\Delta f_n(H) = f_n(H) - f_n(0)$ of HBAR resonance frequencies in the magnetic field [13, 14]. The resonance frequencies itself correspond to the extrema in the frequency response of the transducer's electrical impedance. Thus, the SW excitation and detection are performed electrically by the same piezotransducer. These ADSW establish a spin current $(\mathbf{j}_s)_x$ from YIG into the Pt strip

[21]. The ISHE converts the spin current in the Pt film to a conductivity current (short circuit) or an electrostatic dc field **E**$_{ISHE}$ along the *y*-axis (idle circuit) [22].

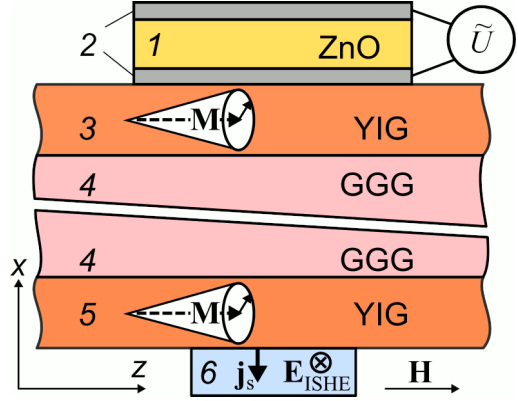

**Fig. 1.** HBAR structure: *1* — ZnO film, *2* —Al electrodes, *3*, *5* —YIG films, *4* — GGG substrate, *6* — thin film Pt strip with direction **a** perpendicular to the figure plane. The typical layer thicknesses: $l^{(1)} = 3$ μm, $l^{(2)} = 200$ nm, $l^{(3,5)} = s = 30$ μm, $l^{(4)} = 500$ μm, $l^{(6)} = 12$ nm. The overlapping area of the top and bottom electrodes *2* has a diameter $a = |\vec{a}| = 170$ μm.

## 3. Theory

Further, we assume that all the thicknesses of the layers $l^{(i)}$ are much smaller than transverse dimensions in the plane (*y*, *z*). In this case, in linear approximation, all variables depend on coordinate *x* and time *t* as $\exp[j(k^{(i)}x - \omega t)]$, where $j = \sqrt{-1}$, $\omega = 2\pi f$, $k^{(i)}$ is a wave number.

For each nonmagnetic layers (*i*=1,4) Newton equation of motion for the elastic displacement $u = u_z$ along with Hooke's law lead to the relationships $k^{(i)} = \omega/V^{(i)}$. Here $V^{(i)} = \sqrt{C^{(i)}/\rho^{(i)}}$ is AW velocity, $\rho^{(i)}$ is the mass density, $C^{(i)}$ is the effective elastic modulus accounting for piezoelectric stiffening in layer *1* [23]. Using two roots $\pm k^{(i)}$, one can obtain the general solutions for $u^{(i)}$. The normal stress can be represented as $T^{(i)} = T^{(i)}_{zx} = C^{(i)}(\partial u^{(i)}/\partial x) + eI\delta_{i1}/(j\omega l^{(1)}C_0)$, where second term exists only in the piezoelectric layer *1*. Here *e* and $C_0$ are the piezoelectric modulus and capacity of the layer, *I* is the displacement current flowing in the layer [24].

For YIG layers (*i*=3, 5) from the Newton equation and Landau-Lifshitz equation for the precession of magnetization vector $\mathbf{M} = (m_x, m_y, M_0)$ together with Maxwell equations, we obtain the secular equation in the form

$$(\omega^2 - k^2V^2)(\omega^2 - \omega_0^2) - \xi\omega_H\omega_M k^2 V^2 = 0, \qquad (1)$$

where we omit the upper indices (*i*). Here, $\omega_0^2 \equiv \omega_0^2(k^2) = \omega_H(k^2)[\omega_H(k^2) + \omega_M]$, $\omega_H(k^2) = \gamma H_{eff}(k^2)$, $\omega_M = 4\pi\gamma M_{eff}$, $H_{eff}(k^2) = H + Dk^2$ and $M_{eff} \approx M_0$ are the uniform effective magnetic field and magnetization, $D$ is the exchange stiffness, $\gamma$ is the gyromagnetic ratio, $\xi = b^2/(4\pi CM_{\text{eff}}^2)$ is the dimensionless coupling parameter, and $b$ is the magnetoelastic constant [15, 25, 26]. Hereinafter, the system of Gaussian-CGS units is used, but for convenience, the layer thicknesses are given in the SI system: micro- and nanometers.

As it is known the crossover of two independent solutions (1) in case of $\xi = 0$ determines MER frequency and wave number: $\omega_{\text{MER}}(H)=2\pi f_{\text{MER}}(H)$ and $k_{\text{MER}}(H)$. In case of $\xi \neq 0$ the formation of coupled waves and the repulsion of the solutions in the vicinity of $\omega_{\text{MER}}$ take place. As one can see, for a given real, positive $\omega$, there are three real roots of the secular equation, $k^2_p(\omega)$ ($p = 1,2,3$). Using six roots $\pm k_{1,2,3}$, one can obtain the general solutions for $u$, $m_{x,y}$, and normal stress component $T = T_{zx} = C(\partial u/\partial x) + bm_x/M_0$.

The solutions obtained for all layers should satisfy the elastic and electrodynamic boundary conditions at the interfaces. At the magnetic layers' interfaces, the additional magnetization boundary conditions for ac magnetization should be taken into account. Here the case of free spins is considered: $\partial m_{x,y}/\partial x = 0$.

Magnetic and acoustic losses are taken into account phenomenologically with the help of the following substitutions: $C^{(i)} \to C^{(i)} + i\omega\eta^{(i)}$ and $\omega_H \to \omega_H + i\gamma\Delta H$, where, $\eta^{(i)}$ and $\Delta H$ are viscosity factor and ferromagnetic resonance (FMR) line width [23, 27].

Further we use the impedance method for calculating the multilayer resonator structure characteristics [23]. The input electric impedance of piezotransducer may be represented as [28]

$$Z_E = \tilde{U}/I = (1 + Z_{\text{AW}})/(j\omega C_0), \qquad (2)$$

where, $Z_{\text{AW}}$ is the function of transducer parameters (material and geometrical) and of the acoustic impedance $Z$ of transducer load (layers *2-6*). It follows from the impedance continuity condition that the load impedance of layer $i$-1 is equal to the input impedance of layer $i$: $z_{\text{in}}^{(i)} = T(x_i + l^{(i)})/(\partial u(x_i + l^{(i)})/\partial t)$. Since the influence of 150–200 nm thickness electrodes on the properties of HBAR is negligible, we can assume that $Z = z_{\text{in}}^{(3)}$. In the absence of magnetoelastic interaction ($\xi = 0$), the load impedance is calculated by the sequential application of the impedance transformation formula [23]

$$z_{\text{in}}^{(i)} = z^{(i)}(z_{\text{in}}^{(i+1)} \cos\varphi^{(i)} + jz^{(i)} \sin\varphi^{(i)})/(z^{(i)} \cos\varphi^{(i)} + jz_{\text{in}}^{(i+1)} \sin\varphi^{(i)}), \qquad (3)$$

where $\varphi^{(i)} = k^{(i)}l^{(i)}$ and $z^{(i)} = \rho^{(i)}V^{(i)}$ are the phase shift and the material acoustic impedance.

For magnetic layers with magnetoelastic interaction, the expression (3) is not applicable, because all three roots $k^2_{1,2,3}$ of the secular equation should be taken into account in the general solution. By matching boundary conditions at YIG surfaces we obtain for input acoustic impedances $z_{\text{in}}^{(3,5)}$ the formula analogous to (3) with the corresponding substitutions:

$$z^{(3,5)} \to \frac{j\sqrt{z_1 z_2}}{\tilde{\omega}}, \quad \sin\varphi^{(3,5)} \to \frac{-2\sqrt{z_1 z_2}}{z_1 + z_2}, \quad \cos\varphi^{(3,5)} \to \frac{z_1 - z_2}{z_1 + z_2}. \qquad (4)$$

Here $z_1 = \sum_{p=1}^{3} \alpha_p \gamma_p \operatorname{tg}(k_p s/2)$, $z_2 = \sum_{p=1}^{3} \alpha_p \gamma_p \operatorname{ctg}(k_p s/2)$, $\tilde{\omega} = \omega \sum_{p=1}^{3} \gamma_p$, $s = l^{(3,5)}$, and $\alpha_p, \gamma_p$ are the coefficients determined in [28]. Thus, the relations (2) - (4) allow us to describe the HBAR spectrum, and determine $(f,H)$ dependences of $f_n$ and resonance $Q_n$ factors.

Let us now consider the features of acoustic spin pumping in our structure. The time-averaged spin current polarized along $\mathbf{z}$, $\mathbf{j}_s \propto g_r \langle (\mathbf{m} \times \partial \mathbf{m}/\partial t)|_{x=x_5} \rangle \propto \theta^2 \mathbf{n}$, flows from YIG layer 5 into Pt layer 6 [21]. Here $g_r$ is the real part of spin mixing conductance, $\theta = \sqrt{\operatorname{Im}[m_x^*(x_5)m_y(x_5)/M_0^2]}$ is the magnetization precession cone angle at YIG/Pt interface $x_5$, and $\mathbf{n}$ is the normal to the interface. The ISHE in Pt leads to an electrostatic field $\mathbf{E}_{\text{ISHE}} \propto -\theta_{\text{SH}}(\mathbf{j}_s \times \mathbf{z}) \propto -\theta^2(\mathbf{n} \times \mathbf{z})$, where $\theta_{\text{SH}}$ is the spin Hall angle of Pt [22]. For a rectangular Pt strip, the dc voltage between its ends in the direction $\mathbf{a}$ is

$$U_{\text{ISHE}} = -(\mathbf{E}_{\text{ISHE}} \cdot \mathbf{a}) \propto \theta^2 ((\mathbf{n} \times \mathbf{z}) \cdot \mathbf{a}). \qquad (5)$$

In (5), the constants $g_r$ and $\theta_{\text{SH}}$ are omitted, since their values are considered to be independent of the field, frequency, and thickness of YIG and Pt. We also omitted the factor resulting from the current density averaging across the Pt thickness.

After substitution the general solutions in magnetic layer 5 to magnetic and elastic boundary conditions we obtain

$$\begin{pmatrix} m_x \\ m_y \end{pmatrix}\bigg|_{x=x_5} = -j\omega \frac{u(x_4)}{\tilde{\omega}(z_1 - z_2)} \sum_{p,q=1}^{3} N_{pq} \begin{pmatrix} \alpha_p \beta_q - \beta_p \alpha_q \\ \alpha_p \delta_q - \delta_p \alpha_q \end{pmatrix}, \qquad (6)$$

where $N_{pq} = \gamma_p \gamma_q [\operatorname{tg}^2(k_p s/2) - \operatorname{tg}^2(k_q s/2)]/[\operatorname{tg}(k_p s/2)\operatorname{tg}(k_q s/2)]$. An explicit view of the amplitude coefficients $\beta_p$ and $\delta_p$ is given in [28].

For allowance of the ADSWs back action on all elastic subsystems, the displacement $u(x_4)$ should be expressed via an electrical parameter of the transducer, for example, the voltage $\tilde{U}$ applied to the electrodes. The transformation

$$u(x_i) = u(x_{i-1})z^{(i)}/(z^{(i)}\cos\varphi^{(i)} + jz_{in}^{(i+1)}\sin\varphi^{(i)}) \qquad (7)$$

is used to express $u(x_4)$ in terms of $u(x_3)$. For the transformation of $u(x_3)$ to $u(x_2)$ via (7) the substitution rules (4) are needed. Finally, from the equations for the piezoelectric layer *1* [24], we obtain

$$u(x_2) = 2je\tilde{U}\sin^2(\varphi^{(1)}/2)/[\omega l^{(1)}(iz^{(1)}\sin\varphi^{(1)} + Z\cos\varphi^{(1)})(1+Z_{AW})]. \qquad (8)$$

Substituting (4), (6) - (8) in (5) we can get an analytical expression for $U_{ISHE}$, which is then analyzed numerically.

## 4. Results and discussions

Figures 2(a),(b) demonstrate the frequency dependences of $Re[k_{1,2,3}(f)]$ and $Im[k_{1,2,3}(f)]$ for infinite magnetic media at $H = 740$ Oe. We attribute $k_1$, $k_2$ to the continuous magnetoelastic branches, which for $f < f_{MER}$ are quasi AW and SW, but for $f > f_{MER}$ are quasi SW and AW. The third root, $k_3(f)$, is always imaginary and plays a certain role for the satisfaction of boundary conditions as well as the root $k_2(f)$ in case $f < f_{FMR} = \omega_0(0)/2\pi$ [4, 29, 30].

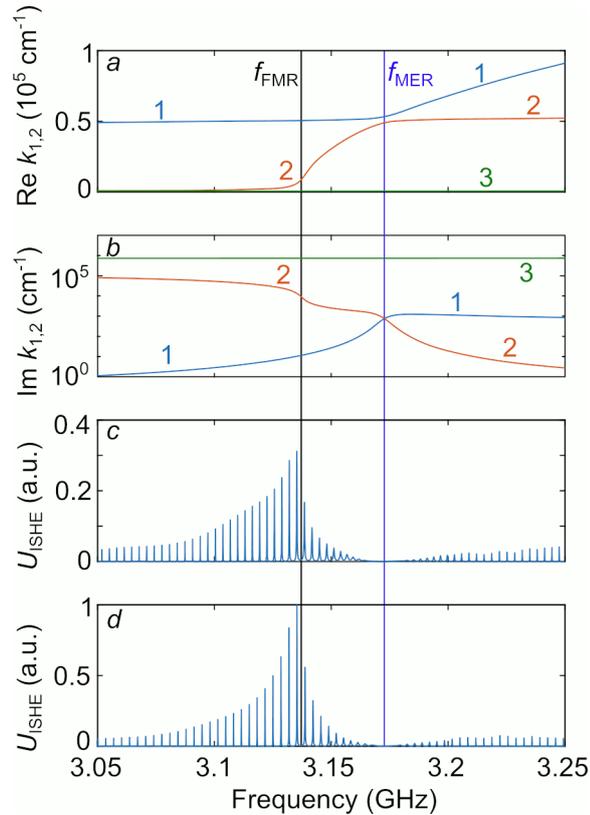

**Fig. 2.** Frequency dependences of: real (a) and imaginary (b) parts of wave numbers $k_1$ (blue lines *1*), $k_2$ (red lines *2*), and $k_3$ (green lines *3*); normalized voltages for structures II (c) and I (d) at fixed magnetic field $H = 740$ Oe. The elastic and magnetic parameters: (111) oriented YIG – $V^{(3,5)} = 3.9 \times 10^5$ cm/s, $\rho^{(3,5)} =$ 5.17 g/cm³ [15], $b = 4 \times 10^6$ erg/cm³, $D = 4.46 \times 10^{-9}$ Oe cm², $4\pi M_{eff} = 955$ G [30], $\Delta H = 0.7$ Oe; GGG –

$V^{(4)} = 3.57 \times 10^5$ cm/s, $\rho^{(4)} = 7.08$ g/cm$^3$; ZnO – $V^{(1)} = 2.88 \times 10^5$ cm/s, $\rho^{(1)} = 5.68$ g/cm$^3$. The layer thicknesses are given in caption to Fig.1

Figures 2 (c), (d) show the frequency dependence of normalized $U_{ISHE}(f)$ at $H= 740$ Oe for two structures, I and II: with two YIG films (II) (Fig.2(c)) and with only one film 5 (I) (Fig.2(d)). The parameters of the YIG / Pt interface and the Pt itself are not considered in this case and the voltage was normalized to the maximum for the structure I. As one can see from Fig. 2(c), (d) the dependences for both structures, I and II are similar in the form and differs only by the scale. Note that the absolute maximum of $U_{ISHE}(f)$ corresponds not to $f_{MER}$, but to a lower frequency near $f_{FMR}$. The local maxima frequencies, as it was shown in [28], coincide with HBAR resonance frequencies.

Next, we consider the frequency dependences in a varying magnetic field and compare them with the experimental ones observed previously in [18]. Figure 3(a) shows for the structure II the calculated dependence $U_{ISHE}(f,H)$, which is in a good agreement with the experimental results, shown in Fig. 3(b). The voltage magnitudes follow the change in the position of the resonance frequencies $f_n(H)$ - the red lines in Fig. 3(a) and the red points in Fig. 3(b). Both calculated and experimental $U_{ISHE}$ magnitudes have different behavior above and below the line $f_{MER}(H)$ (line 1), which is a consequence of excitation of SW waves with basically different wavenumbers. Higher than $f_{MER}(H)$ line, the short SWs with wavenumbers more than $5 \times 10^4$ cm$^{-1}$ are excited, whereas below the line the wavenumbers of excited modes are essentially smaller (see Fig.2(a)). Fitting of the theoretical and experimental dependences allows us to evaluate magnetic parameters $b$, $4\pi M_{eff}$, and $D$ listed in the Fig.2(a) caption [31].

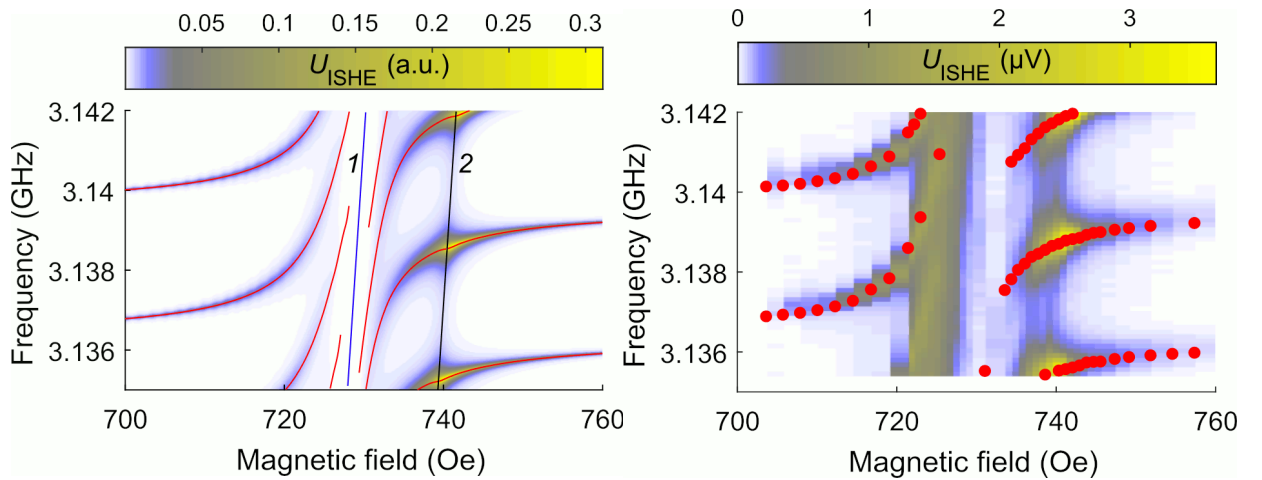

**Fig. 3.** 3D colour plot $U_{ISHE}(f,H)$ for the structure II: (a) – calculated, (b) – experimental adopted from [18] (http://creativecommons.org./licenses/by/4.0/). The red lines (a) and points (b) correspond to positions of HBAR resonant frequency $f_n$ positions. The calculated frequencies $f_{MER}(H)$ and $f_{FMR}(H)$ are shown by line 1 and 2 in (a). For calculation the same parameters as listed in Fig.2 were used. They correspond to the experimental ones.

Up to this point, the consideration concerned rather thick YIG films of about 30 μm. But today, much attention is paid to the technology and the study of submicron and nanometer-sized YIG films. In particular, the effective magnetoelastic interaction in such films was observed [32]. Let us consider the effect of YIG thickness $l^{(5)} = s$ on the $U_{ISHE}$ for the structure I. Figure 4 shows the calculated $U_{ISHE}(f_{0\,max}, s)$ dependence for a maximum located at frequency $f_{0\,max} \approx$ closest to $f_{FMR}$. With the decrease of $s$ from a few tens microns to a micron, the effect increases by an order of magnitude, oscillating with the period ~ 0.65 μm, which corresponds to the AW half-length. In this case, the local minima correspond to $s = (p+1)V/(2f)$ and maxima – to $s = (p+0.5)V/(2f)$, where $p$ is an integer. At $s \sim 3 \div 2$ μm other excitation zones appear at higher frequencies $f_{t\,max}$, corresponding to the SW resonance conditions for free spins: $k_2 s \approx \pi t$, where $t = 1, 2, 3, ...$ A detailed description of the behavior of these high order SW resonance is beyond the scope of this paper. We just note that at certain $s$ the value $U_{ISHE}(f_{t\,max}, s)$ induced by ADSW resonances with $t \leq 3$ becomes larger than the $U_{ISHE}(f_{0\,max}, s)$ (as it is shown in the insertion). With the decrease $s$ up to 120 nm the frequencies $f_{t\,max}$ become so high that are shifted out of the MER influence region. Finally, at the $s \sim 100$ nm only one signal $U_{ISHE}(f_{0\,max}, s)$ remains in the spectrum. With a further thickness decrease, the signal reaches a maximum and then begins to drop.

Since in these calculations we did not take into account the additional magnetic damping due to the spin pumping, the dependence on thickness is entirely determined by the method of the SW excitation. It should be noted that at thicknesses $s < 200$ nm the additional magnetic damping becomes noticeable. Assuming that the FMR linewidth broadening $\Delta H_{sp} \sim g_r/s$ [21, 33] one can obtain a steeper curve for the small thicknesses.

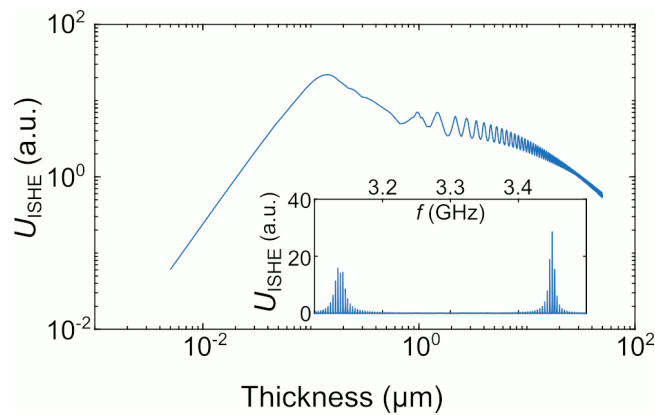

**Fig. 4.** The voltage $U_{ISHE}(f_{0\,max})$ dependence on YIG thickness $s$ at the fixed magnetic field $H = 740$ Oe. The insertion: the spectrum of the signal $U_{ISHE}(f, s=0.2\,\mu m)$ with two zones of maximal excitation near $f_{0\,max}$ and $f_{1\,max}$.

## 5. Conclusion

A semi-analytic theory for ADSW in hypersonic composite HBAR with ZnO-GGG-YIG/Pt layered structure is developed. The theoretical and experimental dependences of the electric voltage $U_{ISHE}(f, H)$ in Pt are in good agreement: the significant asymmetry of the $U_{ISHE}(f_n(H))$ value in reference to the magnetoelastic resonance line $f_{MER}(H)$ position, experimentally observed previously, manifests itself also in the theoretical calculations. This asymmetry is due to the SW spectrum governed by nonuniform exchange: near the $f_{FMR}$ the efficiency of quasiuniform SW excitation is higher than the efficiency for the frequencies exceeding $f_{MER}$. The theory involved takes into account the self-consistent mutual influence of the AW and SW and gives the possibility to evaluate some magnetic parameters of the YIG films including exchange stiffness. The analysis of YIG film thickness influence on $U_{ISHE}$ at the main frequency $f_{0\ max}$ shows that this value reaches a maximum for thicknesses about 100 nm. Also, $U_{ISHE}$ maxima due to the high SW resonances at higher frequencies can be detected. So we believe that acoustic spin pumping created by means of HBAR is a sensitive spectroscopic technique for the investigation of magnetic films properties.

**Acknowledgments**

This work was partially supported by grants 16-07-01210 and 17-07-01498 from the Russian Foundation for Basic Research.

## REFERENCES

[1] A. S. Salasyuk, A. V. Rudkovskaya, A. P. Danilov, B. A. Glavin, S. M. Kukhtaruk, M. Wang, A. W. Rushforth, P. A. Nekludova, S. V. Sokolov, A. A. Elistratov, D. R. Yakovlev, M. Bayer, A. V. Akimov, A. V. Scherbakov, Generation of a localized microwave magnetic field by coherent phonons in a ferromagnetic nanograting, Phys.Rev. B, 97 (2018) 060404(R). https://doi.org/10.1103/PhysRevB.97.060404

[2] P.Graczyk, M. Krawczyk, Coupled-mode theory for the interaction between acoustic waves and spin waves in magnonic-phononic crystals: Propagating magnetoelastic waves, Phys.Rev. B (2017) 024407. https://doi.org/10.1103/PhysRevB.96.024407

[3] A. Barra, A. Mal, G. Carman, A. Sepulveda, Voltage induced mechanical/spin wave propagation over long distances, Appl. Phys. Lett. 110 (2017), 072401. http://dx.doi.org/10.1063/1.4975828

[4] A. Kamra, H. Keshtgar, P. Yan, G. E. W. Bauer, Coherent elastic excitation of spin waves, Phys.Rev. B, 91 (2015) 104409. https://doi.org/10.1103/PhysRevB.91.104409

[5] A. V. Azovtsev, N. A. Pertsev Coupled magnetic and elastic dynamics generated by a shear wave propagating in ferromagnetic heterostructure Appl. Phys. Lett., 111 (2017) 222403. https://doi.org/10.1063/1.5008572

[6] P. Kuszewski, I. S. Camara, N. Biarrotte, L .Becerra, J. von Bardeleben, W. Savero Torres, A. Lemaître , C .Gourdon, J.-Y. Duquesne, L Thevenard. Resonant magneto-acoustic switching: influence of Rayleigh wave frequency and wavevector, J. Phys.: Cond. Matt., 30 (2018) 244003. https://doi.org/10.1088/1361-648X/aac152

[7] K. Uchida, T An., Y. Kajiwara, M. Toda, E. Saitoh. Surface-acoustic-wave-driven spin pumping in Y3Fe5O12/Pt hybrid structure //Appl. Phys. Lett., 99 (2011) 212501. http://dx.doi.org/10.1063/1.3662032

[8] L. Dreher, M. Weiler, M. Pernpeintner, H. Huebl, R. Gross, M. S. Brandt, S. T. B. Goennenwein. Surface acoustic wave driven ferromagnetic resonance in nickel thin films: Theory and experiment. Phys. Rev. B, 86 (2012) 134415. https://doi.org/10.1103/PhysRevB.86.134415

[9] M. Weiler, H. Huebl, F. S. Goerg, F. D. Czeschka, R. Gross, S. T. B. Goennenwein, Spin pumping with coherent elastic waves, Phys. Rev. Lett., 108 (2012) 176601. https://doi.org/10.1103/PhysRevLett.108.176601

[10] X. Li, D. Labanowski, S. Salahuddin, C. S. Lynch, Spin wave generation by surface acoustic waves, J. Appl. Phys., 122 (2017) 043904. http://dx.doi.org/10.1063/1.4996102

[11] S. Cherepov, P. K. Amiri, J. G. Alzate, K. Wong, M. Lewis, P. Upadhyaya, J. Nath, M. Bao, A. Bur, T. Wu, G. P. Carman, A. Khitun, K. L. Wang, Electric-field-induced spin wave generation using multiferroic magnetoelectric cells, Appl. Phys. Lett., 104 (2014) 082403. https://doi.org/10.1063/1.4865916

[12] G. Yu, H. Zhang, Y. Li, J. Li, D. Zhang, N. Sun, Resonance of magnetization excited by voltage in magnetoelectric heterostructures, *Mater. Res. Express,* 5 (2018) 045021. https://doi.org/10.1088/2053-1591/aab91a

[13] N. Polzikova, S. Alekseev, I. Kotelyanskii, A. Raevskiy, Yu. Fetisov, Magnetic field tunable acoustic resonator with ferromagnetic-ferroelectric layered structure, J. Appl. Phys., 113 (2013) 17C704. https://doi.org/10.1063/1.4793774

[14] N. I. Polzikova, A. O. Raevskii, A. S. Goremykina, Calculation of the spectral characteristics of an acoustic resonator containing layered multiferroic structure, J. Commun. Technol. Electron., 58 (2013) 87-94. https://doi.org/10.1134/S1064226912120066

[15] W.Strauss, Magnetoelastic properties of yttrium iron garnet, in W.P. Mason (Ed.), Physical Acoustics, Vol. IV(B), Academic Press, New York, 1968, pp. 211-268.

[16] B. P. Sorokin, G. M. Kvashnin, A. S. Novoselov, V. S. Bormashov, A. V. Golovanov, S. I. Burkov, V. D. Blank, Excitation of hypersonic acoustic waves in diamond-based piezoelectric layered structure on the microwave frequencies up to 20 GHz, Ultrasonics, 78 (2017) 162–165. https://doi.org/10.1016/j.ultras.2017.01.014

[17] N. I. Polzikova, S. G. Alekseev, I. I. Pyataikin, I. M. Kotelyanskii, V. A. Luzanov, A. P. Orlov, Acoustic spin pumping in magnetoelectric bulk acoustic wave resonator, AIP Advances, 6 (2016) 056306. http://dx.doi.org/10.1063/1.4943765

[18] N. I. Polzikova, S. G. Alekseev, I. I. Pyataikin, V. A. Luzanov, A. O. Raevskiy, V. A. Kotov, Frequency and magnetic field mapping of magnetoelastic spin pumping in high overtone bulk acoustic wave resonator, AIP Advances, 8 (2018) 056128. https://doi.org/10.1063/1.5007685

[19] Note that the presence of the upper YIG film *3* is not necessary. However, we take into consideration this film because such a structure we used in the experiments [18].

[20] V. A. Luzanov, S. G. Alekseev, N. I. Polzikova, Deposition process optimization of zinc oxide films with inclined texture axis, J. Commun. Technol. Electron., 63 (2018) 1076–1079. DOI:10.1134/S1064226918090127

[21] Y. Tserkovnyak, A. Brataas, G.E.W. Bauer, Enhanced Gilbert damping in thin ferromagnetic films, Phys. Rev. Lett., 88 (2002) 117601. https://doi.org/10.1103/PhysRevLett.88.117601

[22] E. Saitoh, M. Ueda, H. Miyajima, G. Tatara, Conversion of spin current into charge current at room temperature: inverse spin-Hall effect, Appl. Phys. Lett., 88 (2006) 182509. https://doi.org/10.1063/1.2199473

[23] B. A. Auld, "Acoustic Fields and Waves in Solids," Vol. I, Wiley, New York, 1973, 423 p.

[24] D. Royer, E. Dieulesaint, Elastic Waves in Solids II. Generation. Acousto-optic Interaction, Applications, Springer-Verlag, Berlin Heidelberg, 2000.

[25] In the geometry involved, the demagnetization field $\mathbf{h}_d = (-4\pi m_x, 0, 0)$ contributes only to the ac component of the effective field. The effect of the YIG crystalline anisotropy is not considered in detail in this work, but it can be taken into account both in $M_{eff}$ and $H_{eff}$. For (111) YIG films orientation, as in Ref. 18, anisotropy contributes mainly to the $M_{eff}$, rather than to the $H_{eff}$ [26]. In this case the coupling constant $b$ is determined by a linear combination of two cubic constants $b_2$ and $b_1$ with coefficients depending on the field direction in the (111) plane [15].

[26] S. Klingler, A. V. Chumak, T. Mewes, B. Khodadadi, C. Mewes, C. Dubs, O. Surzhenko, B. Hillebrands, and A. Conca, Measurements of the exchange stiffness of YIG films using broadband ferromagnetic resonance techniques Journal of Physics D: Applied Physics **48**, (2015) 015001 https://doi.org/10.1088/0022-3727/48/1/015001

[27] A. G. Gurevich and G. A. Melkov, Magnetization Oscillations and Waves (CRC-Press, Boca Raton, 1996), p. 464.

[28] N.I.Polzikova, S.U.Alekseev, V.A.Luzanov, A.O.Raevskiy, Electroacoustic excitation of spin waves and their detection via inverse spin Hall effect, Phys. Solid State, 60 (2018) 2211. https://doi.org/10.1134/S1063783418110252

[29] N.M.Salanskii, M.S.Yerukhimov, The Physical Properties and Applications of Magnetic Films, Nauka, Novosibirsk, 1975 (in Russian).

[30] H.F.Tiersten, Thickness vibrations of saturated magnetoelastic plates, J. Appl. Phys., 36 (1965) 2250-2259. https://doi.org/10.1063/1.1714459

[31] Note that the reduced magnetization $4\pi M_{eff} = 955$ G is characteristic for La, Ga-substituted YIG epitaxial films used in the experiment.

[32] Yu. V. Khivintsev, V. K. Sakharov, S. L. Vysotskii, Yu. A. Filimonov, A. I. Stognii, S. A. Nikitov, Magnetoelastic waves in submicron yttrium–iron garnet films manufactured by means of ion-beam sputtering onto gadolinium–gallium garnet substrates, Technical Phys., 63 (2018) 1029-1035. https://doi.org/10.1134/S1063784218070162

[33] M. B. Jungfleisch, A. V. Chumak, A. Kehlberger, V. Lauer, D. H. Kim, M. C. Onbasli, C. A. Ross, M. Kläui, and B. Hillebrands Thickness and power dependence of the spin-pumping effect in Y3Fe5O12/Pt heterostructures measured by the inverse spin Hall effect, Phys. Rev. B 91, (2015)  134407 https://doi.org/10.1103/PhysRevB.91.134407